\documentclass[twocolumn,floatfix,longbibliography,hyperref, prl]{revtex4-1}
\usepackage{amsfonts,amsmath,graphicx,amssymb,bm}
\usepackage[pdftex,plainpages=false,colorlinks=true,linkcolor=blue, citecolor=blue, urlcolor=blue]{hyperref}
\pdfoutput=2

\usepackage[charter]{mathdesign}
\usepackage[x11names,svgnames]{xcolor}

\newcommand\Sigmav{\bm{\Sigma}}

\newcommand\Gv{\mathbf{G}}
\newcommand\Tr{\mathrm{Tr}}

\begin{document}

\title{Pseudogap to metal transition in the anisotropic two-dimensional Hubbard model}
\author{J. P. L. Faye}
\affiliation{The Abdus Salam International Center for Theoretical Physics, Strada Costiera 11, 34014 Trieste, Italy}
\author{D. S\'en\'echal}
\affiliation{D\'epartement de physique and Institut Quantique, Universit\'e de Sherbrooke, Sherbrooke, Qu\'ebec, Canada J1K 2R1}
\date{\today}

\begin{abstract}
The transition between a metallic and  a pseudogap phase in high-$T_c$ cuprate superconductors is the subject of experimental investigations but has not been settled theoretically, even within the context of the Hubbard model.  
We apply the Cluster Dynamical Impurity Approximation (CDIA) to the anisotropic Hubbard model on the square lattice at zero temperature and finite doping.
This approach can detect a first-order transition between two metallic states: a pseudogap state at low doping, with a depleted density of states at the Fermi level, and a correlated metal at higher doping.
This transition was first seen in Cluster dynamical mean field theory at finite temperature by Sordi et al on the isotropic Hubbard model.
Here we investigate this transition at zero temperature and a as function of on-site interaction $U$, anisotropy $t_y/t_x$ and doping.
We find a first-order transition line which ends at a quantum critical point, which occurs around $t_y/t_x\sim 0.5$.
\end{abstract}
\maketitle

The pseudogap phenomenon is one of the main experimental signatures of strongly correlated physics in hole-doped, high-temperature superconductors.
It can be defined as a loss of density of states at the Fermi level as one lowers the temperature below a crossover temperature $T^*$, or at very low temperatures as one moves towards half-filling from the overdoped region of the phase diagram, below a doping level $p^*$.
The phenomenon can be seen in NMR experiments, in STM spectroscopy and in ARPES (for a review, see Ref.~\cite{Norman2005}).
Recently, a sharp drop in the Hall number when doping is decreased below $p^*$ has been observed in YBCO\cite{Badoux2016}.
This drop is interpreted as a sudden loss of carriers across some sort of zero-temperature pseudogap transition.
Phenomenological models have been proposed in order to describe this transition~\cite{Storey2016a, Eberlein:2016, Yang2006a, Chatterjee2016, Maharaj2016, Morice2017, Verret2017}.
Many of these models see the drop as a result of a reconnection of the Fermi surface at a Lifshitz transition caused by long-range order, for instance antiferromagnetism.

A possible alternative explanation, to which the present work lends some support, is that the pseudogap transition is not akin to a Lifshitz transition caused by the onset of long range order, but rather to a first-order transition between a metallic phase and a strongly correlated, pseudogap phase.
In the pseudogap phase, the effective number of Hall carriers would be proportional to the hole doping $p$, whereas in the metallic phase it would behave like the total number $1+p$ of electrons. 
The drop in Hall number would then occur within a coexistence region between the two phases.

Sordi et al.~\cite{Sordi2011,Sordi:2012kq}, using cluster dynamical mean field theory (CDMFT) with a continuous-time quantum Monte Carlo impurity solver, have revealed the existence of a first-order transition between a pseudogap phase and a metallic phase at finite doping.
This transition can be seen as an extension of the Mott transition, which occurs at half-filling, to finite doping, and ends with a critical point upon increasing the temperature at a fixed value of the chemical potential.
Such a finite-doping transition was also seen with the dynamical cluster approximation (DCA) at finite temperature~\cite{Macridin2006} and with ED-CDMFT at very low temperature~\cite{Liebsch:2009bv}.
Since the two phases have the same symmetry, a second-order transition between the two should normally only occur at the end of a first-order line.
Beyond this critical point, i.e., at higher temperatures, the first order transition is replaced by a smooth crossover along what is known as a Widom line.
These high-temperature crossovers are also seen on larger clusters~\cite{Gull2010}.

In this work, we will investigate the pseudogap phenomenon at zero temperature in the anisotropic, square lattice Hubbard model:
\begin{multline}\label{eq:Hubbard}
H = -\sum_{i,\sigma}\left\{ t_x\, c^\dagger_{i,\sigma} c_{i+x,\sigma} + t_y\, c^\dagger_{i,\sigma} c_{i+y,\sigma}\right\} 
 + \mathrm{H.c.} \\  
+ U \sum_i n_{i \uparrow} n_{i \downarrow} - \mu \sum_{i, \sigma} n_{i \sigma}\, .
\end{multline}
where $t_x$ and $t_y$ are the hopping amplitudes in the $x$ and $y$ directions, $U$ is the on-site interaction, $n_{i\sigma}=c^\dagger_{i\sigma} c_{i\sigma}$ is the number of electrons of spin $\sigma=\uparrow,\downarrow$ at site $i$ and $\mu$ is the chemical potential.
We use a method very close to dynamical mean field theory: The cluster dynamical impurity approximation (CDIA).
The motivation behind studying the spatially anisotropic Hubbard model is partly methodological:
It could be argued that the first-order transition (and hence the well-defined pseudogap state) seen in Ref.~\cite{Sordi2011,Sordi:2012kq} is an artefact of the method used, i.e., dynamical mean field theory on a relatively small cluster (4 sites).
If this were the case, one would expect to see such a transition in the one-dimensional model as well, where one knows from the Bethe ansatz solution that none exists~\cite{Lieb:1968fk}.
If the Sordi transition is real, one then expects a first-order line in the $(t_y,\mu)$ plane that ends with a critical point at an intermediate value of $t_y$ ($\mu$ is the chemical potential).
This is indeed what we can infer from our data, with a quantum critical point lying around $t_y=0.5$, depending on $U$ (we will set $t_x=1$ throughout).

Note that the simple square lattice dispersion used here does not allow for a precise modeling of high-$T_c$ cuprates.
Extending our results to dispersions specific to high-$T_c$ materials will be the object of future work.

\begin{figure}
\centerline{\includegraphics[scale=1]{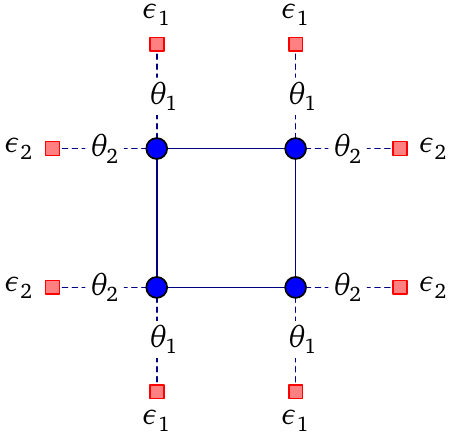}}
\caption{Cluster-bath system used in this work. 
Blue dots represent the cluster sites and red squares the auxiliary (or bath) orbitals forming the effective medium.
$\theta_{1,2}$ are hybridizations amplitudes between site and bath orbitals, and $\epsilon_{1,2}$ are the 
energies of the bath orbitals. In this simple parametrization, symmetry considerations reduce the number of independent bath parameters to four.}
\label{fig:bath}
\end{figure}

Let us first quickly review the methodology used.
The CDIA is based on Potthoff's self-energy functional approach~\cite{Potthoff2003b, Potthoff:2014rt}.
It is very close to CDMFT, in particular its zero-temperature formulation in terms of bath orbitals.
In CDMFT, the Hubbard model on the infinite lattice is replaced by the same model defined on a small
cluster (e.g. a 4-site plaquette) embedded in an effective medium; this is called the {\em impurity model}.
The effective medium is determined self-consistently by imposing the condition that the electron Green function
of the impurity model coincides with the local Green function of the infinite lattice when the self-energy of the
latter is replaced by that of the impurity. 
The impurity problem may be solved numerically, for instance in the so-called hybridization expansion scheme of continuous-time quantum Monte Carlo (CTQMC-HYB), or at zero temperature using exact diagonalization (CDMFT-ED).
In the latter case, the effective medium is represented by a finite number of auxiliary orbitals hybridized with the cluster orbitals (see fig.~\ref{fig:bath}).
The impurity model is then defined by the following Hamiltonian:
\begin{equation}\label{eq:impurity}
H_{\rm imp} = H_{\rm Hub} + \left(\sum_{i,r} \theta_{ir}c^\dagger_{i\sigma}a_{r\sigma} + \mathrm{H.c.}\right) + \sum_r \epsilon_r a_{r\sigma}^\dagger a_{r\sigma}
\end{equation}
where $H_{\rm Hub}$ is the restriction to the cluster of the Hubbard model \eqref{eq:Hubbard}, $a_r$ annihilates an electron on the bath orbital
labeled $r$, $\theta_r$ is a hybridization parameter and $\epsilon_r$ a bath energy.
Practical numerical constraints on the size of the Hilbert space restrict the total number of orbitals (site + bath + spin) to 24, which still amounts to $24!/(12!)^2\sim 2.7\times10^6$ states even when we restrict ourselves to total spin zero.
In this work we adopt the system illustrated in Fig.~\ref{fig:bath}, with a cluster of four sites, plus eight bath orbitals.
Symmetry considerations only leave four independent bath parameters ($\theta_{1,2}$ and $\epsilon_{1,2}$) instead of 16.
In CDIA, the bath parameters defined on Fig.~\ref{fig:bath} are not determined by an {\it ad hoc} self-consistency requirement, but instead by the minima (or saddle points) of the Potthoff functional:
\begin{multline}\label{eq:omega}
 \Omega[\Sigmav(\theta,\epsilon)]=\Omega'[\Sigmav(\theta,\epsilon)]\\ +\Tr\ln[-(\Gv^{-1}_0 -\Sigmav(\theta,\epsilon))^{-1}]
 -\Tr\ln(-\Gv'(\theta,\epsilon))
\end{multline}
where $\theta,\epsilon$ represent the bath parameters, $\Gv'$ is the Green function of the impurity problem computed in exact diagonalization, $\Sigmav$ is the corresponding self-energy, $\Gv_0$ the noninteracting Green function of the infinite-lattice model and $\Omega'$ the ground state free energy ($E-\mu N$) of the impurity problem. The symbol $\Tr$ stands for a functional trace, which amounts to an integral over frequencies along the imaginary axis (once the integrand is properly regularized), an integral over wave vectors and an ordinary trace over spin and band indices (if any).
In practice, one must compute the functional \eqref{eq:omega} numerically for each set $(\theta_i,\epsilon_i)$ of bath parameters, and use an algorithm, such as the Newton-Raphson method, to find the saddle points of $\Omega$. 
This is a costly and delicate numerical task, which explains why the method is not widely applied.
In Ref.~\cite{Balzer:2009kl} the CDIA was used to track the Mott transition in the half-filled, particle-hole symmetric Hubbard model, in which $\theta_2=\theta_1$ and $\epsilon_2=-\epsilon_1$.
In Ref.~\cite{Lenz2016}, the same method was used to investigate the evolution of the Mott transition at half-filling as a function of $t_y$ in the anisotropic Hubbard model \footnote{The variational cluster approximation (VCA) is essentially the same method, except that the bath is nonexistent: Weiss fields defined on the cluster serve as variational parameters instead.};
see also Ref.~\cite{Raczkowski2012}, where finite-temperature CDMFT was applied to the same problem, and Ref.~\cite{Biermann:2001fk} where a variant of DMFT was used.
More explanations on this method and its merits can be found in Ref.~\cite{Senechal:2010fk}.

\begin{figure}
\centerline{\includegraphics[scale=0.8]{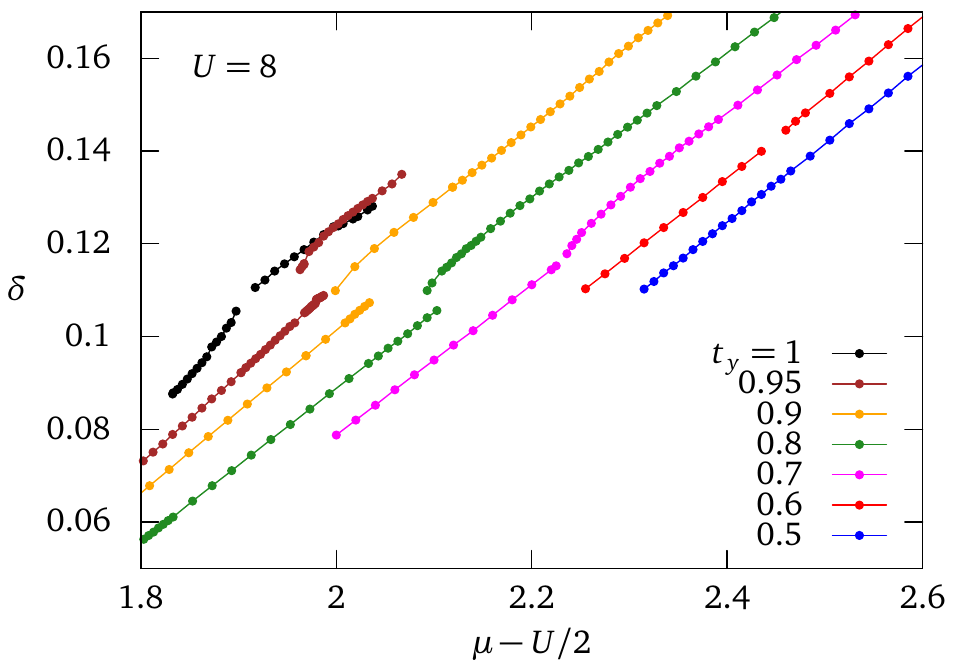}}
\caption{Doping as a function of chemical potential for several values of $t_y$ at $U=8$.}
\label{fig:n_vs_mu}
\end{figure}

\begin{figure}
\centerline{\includegraphics[scale=0.75]{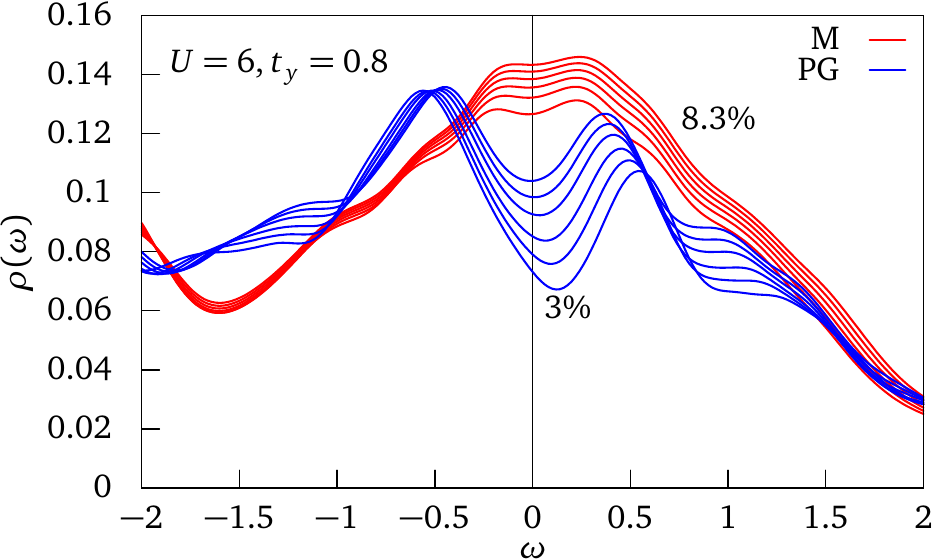}}
\caption{Density of states $\rho(\omega)$ for a few values of doping for $U=6$ and $t_y=0.8$. Blue curves (ranging from 3\% to 6.4\%) cover the pseudogap phase, and red curves (from 5.9\% to 8.3\%) the metallic phase.
A Lorentzian broadening $\eta=0.2$ was used.}
\label{fig:dos}
\end{figure}

We have applied this method to the Hubbard model \eqref{eq:Hubbard} for several values of $t_y$ and $U$, scanning over the chemical potential.
Once a solution, i.e., a converged value of the bath parameters $(\theta_{1,2},\epsilon_{1,2})$, is found, then the associated value of the Potthoff functional $\Omega$ is a good approximation to the infinite system's free energy $E-\mu n$ ($E$ is the energy density and $n$ the electron density).

\begin{figure}
\centerline{\includegraphics[scale=0.8]{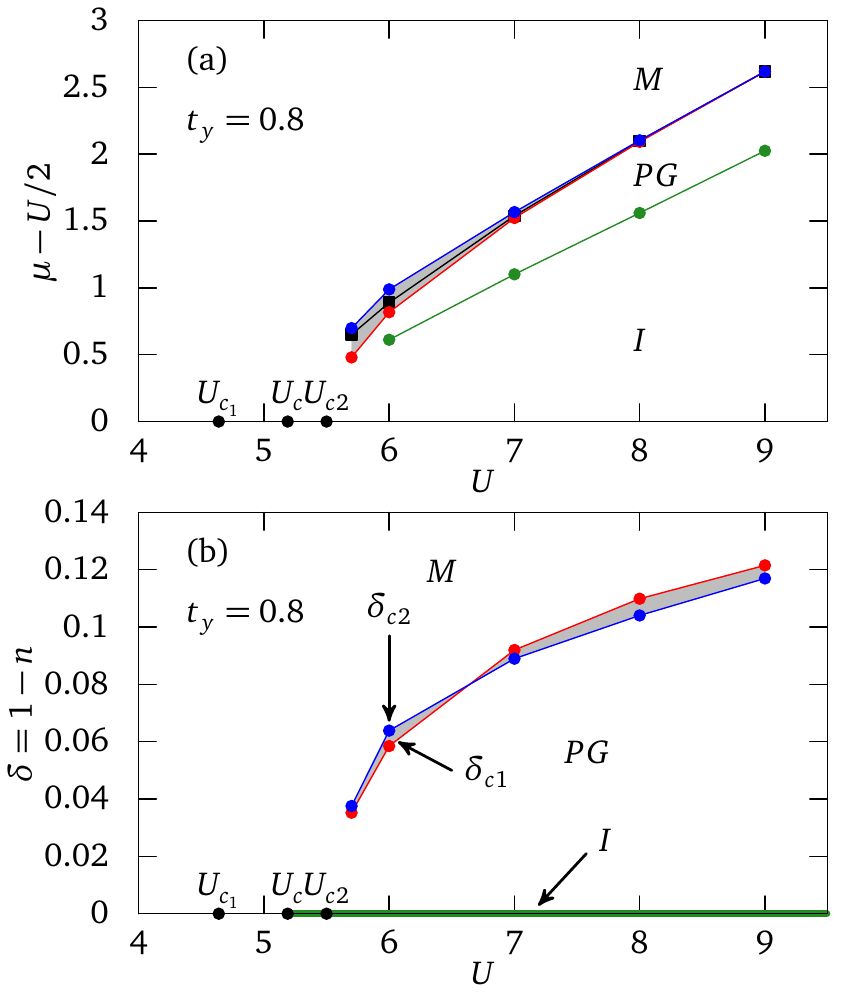}}
\caption{Location of the pseudogap (PG) to metal (M) transition in the $U-\mu$ plane (a) and the $U-\delta$ plane (b), for $t_y=0.8$. The black squares in (a) indicate the point where the energies of the two solutions cross. The blue dots indicate the last metastable pseudogap solution (coming from half-filling) and the red dots the last metastable metallic solution (coming towards half-filling). The green dots (top panel) and the green line (bottom panel) indicate the boundary of the Mott phase (I), at half-filling.}
\label{fig:phase_diagram_U-mu}
\end{figure}

\begin{figure}
\centerline{\includegraphics[scale=0.8]{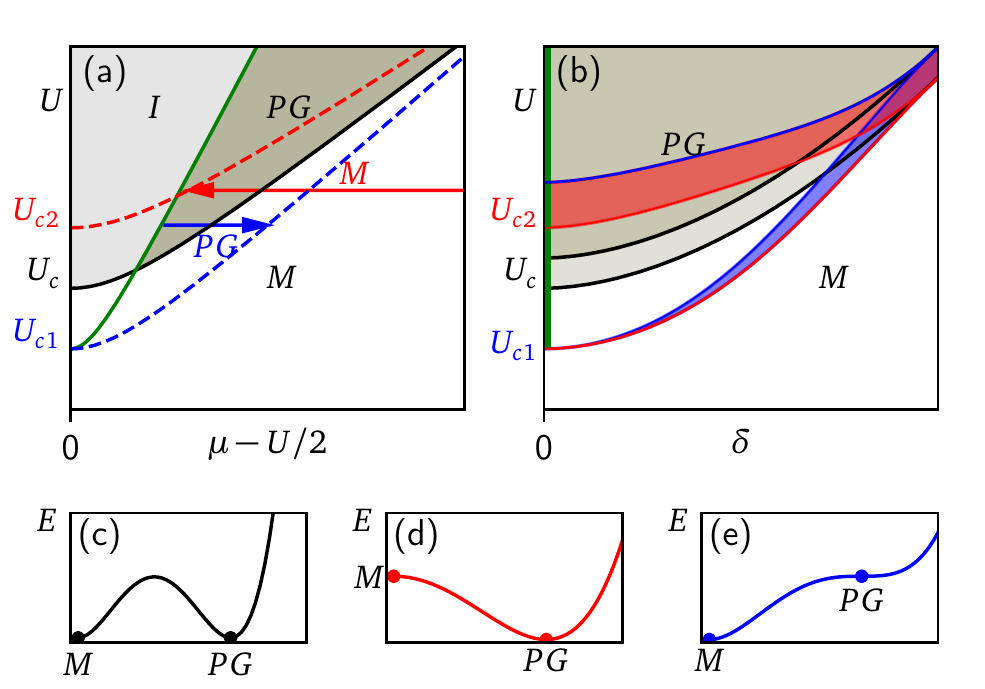}}
\caption{Scenario for a phase diagram at fixed $t_y$ in the (a) $\mu-U$ and (b) $\delta-U$ planes, based on our observations and continuity. In panel (a), the red-dashed curve is the location of the last metastable metallic solution (free energy diagram on panel (d)) and the dashed blue curve the location of the last metastable pseudogap solution (free energy diagram on panel (e)).
The free energies of the pseudogap (PG) and metallic (M) phases cross at the black curve (panel (c)).
The green curve is the location of the second-order transition between the Mott insulator (I) and the pseudogap (PG).
On panel (b), the blue, red and black curves become regions of multiple densities.
}
\label{fig:phase_scenario}
\end{figure}

Fig.~\ref{fig:n_vs_mu} shows the density $n$ as a function of chemical potential for several values of $t_y$, at $U=8$.
The first-order transition is clearly visible, except at $t_y=0.5$ and $t_y=0.6$. 
However, it still exists at $t_y=0.6$, but not at $t_y=0.5$, as can be seen from the jump in bath parameters~\cite{supplement}.
The transition seems weaker close to the isotropic point ($t_y=1$), which is also the most difficult to converge numerically.

Fig.~\ref{fig:dos} shows the density of states $\rho(\omega)$ as a function of frequency for values of doping around the transition, showing a drop of $\rho(0)$ across the transition, into the pseudogap phase.
Fig.~\ref{fig:phase_diagram_U-mu} shows the location of the pseudogap transition for several values of $U$ at $t_y=0.8$.
The top panel shows the computed phase diagram in the $U-\mu$ plane, where the location of the insulating phase at half-filling is also indicated.
The bottom panel shows the same data in the $U-\delta$ plane;
note that the red and blue curves cross at some point on panel~(b). The phase coexistence region is colored in gray.
The location of the Mott transition at half-filling ($U_{c1}$, $U_c$ and $U_{c2}$) is also indicated along the horizontal axis (half-filling).
One of the advantages of CDIA is its ability to provide an estimate of the grand potential, and therefore of the value $U_c$ at which the energies of the insulating and metallic phases cross, and likewise for the pseudogap and metallic phases.
The black squares on the top panel indicate precisely this. 
They should connect to $U_c$  at the particle-hole symmetric point $\mu=U/2$.

Fig.~\ref{fig:phase_scenario}a illustrates a possible phase diagram scenario in the $\mu-U$ plane (note that the two axes are interchanged on this plot compared to the previous one).
Two distinct solutions of the CDIA exist as $\mu$ is scanned: A metallic solution (M) when scanning towards half-filling (red arrow) until the red dashed curve is met, and a pseudogap solution when scanning from half-filling (blue arrow) until the dashed blue curve is met.
The energies of the two solutions cross at the continuous black curve.
The transition between the insulator (I) and the pseudogap (PG) is continuous (i.e. second order) and occurs along the green curve, whereas the transition between the latter and the metallic solution is of first order, characterized by hysteresis between the two dashed curves.
The region of stability of the insulator and pseudogap states are shaded.
Fig.~\ref{fig:phase_scenario}b shows the corresponding phase diagram in the $U-\delta$ plane. 
On this plane each of the red, blue and black curves translates into two curves: One for the PG, the other for the M solution.
Phase coexistence occurs within each pair of curves~\cite{supplement}.

\begin{figure}
\centerline{\includegraphics[scale=0.8]{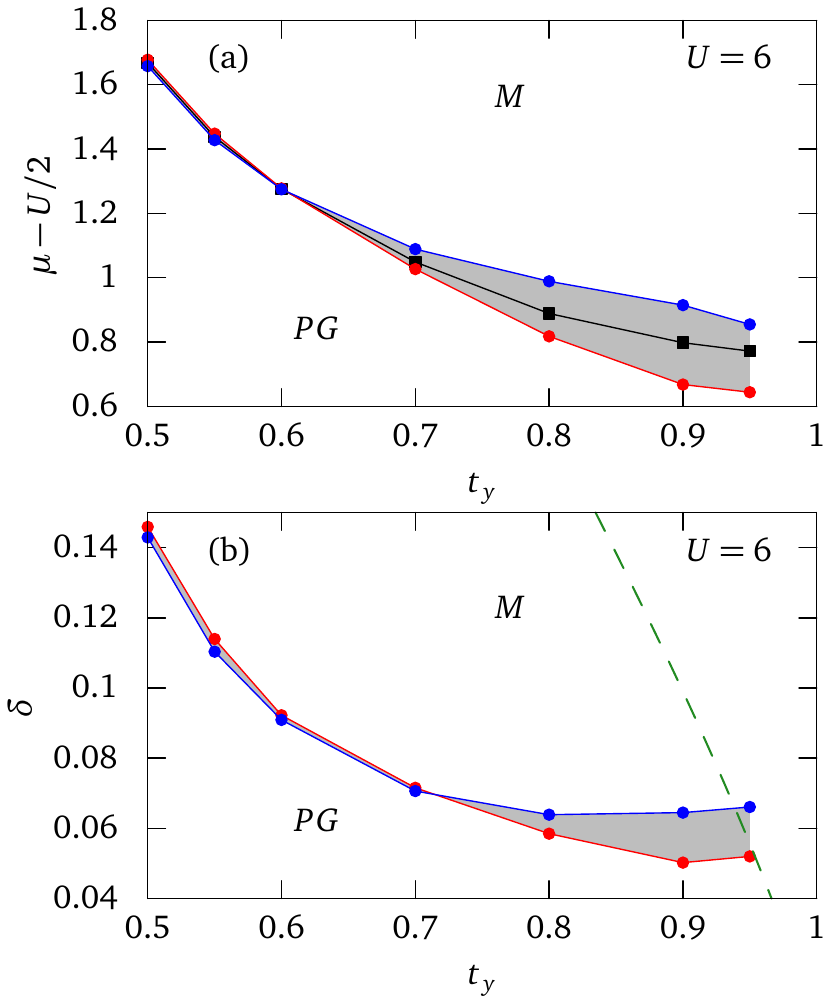}}
\caption{Location of the pseudogap (PG) to metal (M) transition in the $t_y-\mu$ plane (a) and the $t_y-\delta$ plane (b), for $U=6$. In (a), the black squares (mostly hidden) indicate the point where the free energies of the two solutions cross. The blue dots indicate the last metastable pseudogap solution (coming from half-filling) and the red dots the last metastable metallic solution (coming towards half-filling).
The first order transition disappears somewhere around $t_y=0.5$.
The green dashed line indicates the location of the Lifshitz transition at $U=0$.}
\label{fig:phase_diagram_U8}
\end{figure}

Figure~\ref{fig:phase_diagram_U8} shows the computed phase diagram in the $t_y-\mu$ (a) and $t_y-\delta$ (b) planes at $U=6$.
The free energy crossing curve (black squares) is shown on panel (a).
Again, a crossing of the two instability curves (red and blue) occurs on Panel (b).
Hysteresis is practically impossible to see below $t_y=0.6$ and the transition has all but disappeared at $t_y=0.5$ (it has completely disappeared
at $t_y=0.5$, $U=8$~\cite{supplement}).
We conclude that the first-order pseudogap to metal transition must end at a quantum critical point around $t_y=0.5$ for the values of $U$ studied.
This is compatible with the absence of such a transition in the one-dimensional limit ($t_y=0$).
As a function of density, the coexistence region is largest near $t_y=1$.

Some methodological questions can be raised concerning our results.
For instance, when using a finite-size bath, the number $N$ of particles in the system (cluster $+$ bath) is conserved, although the number of particles on the cluster {\it per se} is not. 
When changing the chemical potential, not only does the fraction of $N$ located in the cluster changes, but at some point the ground state of the system could shift from one value of $N$ to another (what we call a sector shift).
In the above calculations, $N$ was always equal to 12 and we checked that the solutions found indeed correspond to the system's ground state. 
Thus, the pseudogap transition cannot be attributed to some change in $N$. 
Such a change is not expected anyway when close to half-filling, and the pseudogap transition is clearly visible near 6\% doping when $U=6$.
Besides, this issue does not arise in the results of Ref.~\cite{Sordi2011,Sordi:2012kq}, where the bath is effectively infinite.

A second issue is the presence of antiferromagnetic (AF) order.
In this work this phase was not probed at all, and we know for a fact that the antiferromagnetic transition will pre-empt the Mott transition in the half-filled Hubbard model as $U$ is increased. 
Therefore, the phase diagrams of Figs~\ref{fig:phase_diagram_U-mu} and \ref{fig:phase_diagram_U8} should also incorporate an AF phase that covers part of the pseudogap transition. 
This phase is not expected to extend too far as a function of doping $\delta$, but how far exactly we cannot say in the precise context of our results.
Probing that phase increases considerably the challenge of our computations since the number of independent bath parameters required to describe it effectively doubles.
If a frustration-inducing diagonal hopping $t'<0$ were added, the extent of the AF phase would be reduced as a function of hole doping, and the Mott state would be revealed with sufficiently high $t'$.

One might ask whether the pseudogap transition observed here is the consequence of an underlying van Hove singularity (or Lifshitz transition).
In the noninteracting, anisotropic Hubbard model, a Lifshitz transition occurs as a function of doping and $t_y$ when the Fermi surface changes topology from closed to open.
This location of this transition is indicated by a green dashed line on Fig.~\ref{fig:phase_diagram_U8}b and has clearly nothing to do with the pseudogap transition we observe.

In conclusion, our CDIA computations support the existence of a zero-temperature, first-order transition at finite doping between a pseudogap and a metallic phases. 
In the $t_y-\delta$ plane, this first-order line ends with a quantum critical point near $t_y=0.5$, which is natural given that the pseudogap does not exist in the one-dimensional limit.
This transition could be the origin of the sudden drop in carrier density observed in transport measurements.

We gratefully acknowledge conversations with A.M.~Tremblay, G.~Sordi and M.~Kiselev.
Computing resources were provided by Compute Canada and Calcul Qu\'ebec.
This research is supported by NSERC grant no RGPIN-2015-05598 (Canada) and by FRQNT (Qu\'ebec).


%

\end{document}